\documentclass[conference,a4paper]{APSIPA2026}

\usepackage{amsmath}
\usepackage{amsfonts}
\usepackage{graphicx}
\usepackage{multirow}

\usepackage{colortbl}
\usepackage[dvipsnames]{xcolor}

\usepackage{url}
\usepackage{breakurl}
\usepackage[breaklinks]{hyperref}

\usepackage{tabularx}
\usepackage{array}
\usepackage{booktabs}
\usepackage[flushleft]{threeparttable}

\newcolumntype{Y}{>{\raggedright\arraybackslash}X}

\usepackage[
  backend=biber,
  style=ieee
]{biblatex}
\usepackage{geometry}
\usepackage{fancyhdr}

\fancypagestyle{firststyle}{
  \fancyhf{}
  \fancyhead[C]{2026 Asia Pacific Signal and Information Processing Association Annual Summit and Conference (APSIPA ASC)}
}

\begin{document}

\title{Closing the Affective Loop: Multimodal Speaker--Listener Emotion-Dynamics-Aware Empathetic Social Robots}

\author{
\authorblockN{Zi Haur Pang\authorrefmark{1},
Casey Kennington\authorrefmark{2}, and
Tatsuya Kawahara\authorrefmark{1}}

\authorblockA{\authorrefmark{1}
Graduate School of Informatics, Kyoto University, Japan\\
E-mail: \{pang,kawahara\}@sap.ist.i.kyoto-u.ac.jp}

\authorblockA{\authorrefmark{2}
Computer Science Department, Boise State University, United States\\
E-mail: caseykennington@boisestate.edu}
}

\maketitle
\thispagestyle{firststyle}
\pagestyle{empty}

\begin{abstract}

Empathetic social robots should respond not only to what users say, but also to
how their emotions dynamically evolve during interaction. However, existing
empathetic dialogue systems are often text-centered and primarily model empathy
as a one-way mapping from the user's emotion to the system response, limiting
their ability to capture embodied speaker--listener affective exchange. We
present \textsc{AffectLoop}, a multimodal speaker--listener
emotion-dynamics-aware spoken dialogue system implemented on the Misty II
robot. The system tracks the speaker's verbal and facial affective dynamics,
estimates the robot listener's own verbal and behavioral affective state, and
conditions LLM-based response generation on both affective streams. The robot
then generates a short spoken empathetic response together with emotionally
congruent embodied behavior, forming a closed speaker--listener affective loop.
We evaluate the system in a pilot within-subject study with five participants, comparing it with an otherwise identical utterance-conditioned baseline that omits the speaker- and listener-affective-state inputs. The proposed system received higher
overall impression ratings, especially for empathetic response
and user satisfaction. Post-hoc log analysis further showed higher
speaker--listener affective alignment and stronger valence-based distress
recovery. These preliminary results suggest that explicitly modeling both
speaker emotional dynamics and listener affective state can improve embodied
empathetic interaction.

\end{abstract}

\section{Introduction}


Empathetic dialogue systems aim to understand a user's emotional situation and generate supportive responses accordingly. Early benchmarks such as \textsc{EmpatheticDialogues} have shown that emotion-based training improves perceived empathy~\cite{rashkin2019towards}, while later models further incorporate emotion mimicry, psychotherapy--based, and emotion flow modeling~\cite{majumder2020mime,pang2024acknowledgment}. However, most existing approaches remain text-centered and primarily model empathy as a one-way mapping from the speaker's emotion to the system's response~\cite{raamkumar2022empathetic}.

This limitation is especially important for social robots, where empathy is conveyed not only through language but also through nonverbal behavior. Recent work has explored using large language models (LLMs) to generate affective robot behaviors, including speech style, gestures, facial expressions, and emotional displays~\cite{lee2023safe,mishra2023realtime}. Nevertheless, these systems rarely model the dynamic affective exchange between the human speaker and the robot listener. In human communication, emotions change over time~\cite{hipson2021emotion}, are expressed through multimodal signals~\cite{sauter2017nonverbal}, and can shape speaker--listener alignment~\cite{smirnov2019emotions}. This suggests that an empathetic robot should consider both the speaker's emotional dynamics and its own affective state as the listener.

To address these limitations, we present \textsc{AffectLoop}, a multimodal speaker--listener emotion-dynamics-aware spoken dialogue system for an empathetic social robot. The system conditions LLM-based response generation on two coupled affective streams: the speaker's verbal and facial emotional dynamics, and the listener's own verbal and behavioral affective state. Based on these signals, the robot generates a short spoken empathetic response together with emotionally congruent embodied behavior.

We conduct a pilot within-subject user study comparing our system with a text-only LLM baseline. Results suggest that the proposed system improves perceived naturalness, empathetic listening, empathetic response, and user satisfaction. A post-hoc interaction-log analysis further shows higher speaker--listener affective alignment and greater distress recovery in the proposed system, suggesting that explicitly closing the affective loop can improve embodied empathetic interaction.

Our contributions are:
\begin{itemize}
    \item We introduce a speaker--listener affective conditioning framework that incorporates both speaker emotional dynamics and listener affective state into LLM-based response generation, deployed in a empathetic social robot.
    \item We provide preliminary user-study and interaction-log evidence showing improvements in perceived empathy, user satisfaction, affective alignment, and distress recovery.
\end{itemize}

\section{Related Work}




\begin{figure*}[ht]
  \centering
  \includegraphics[width=0.79\linewidth]{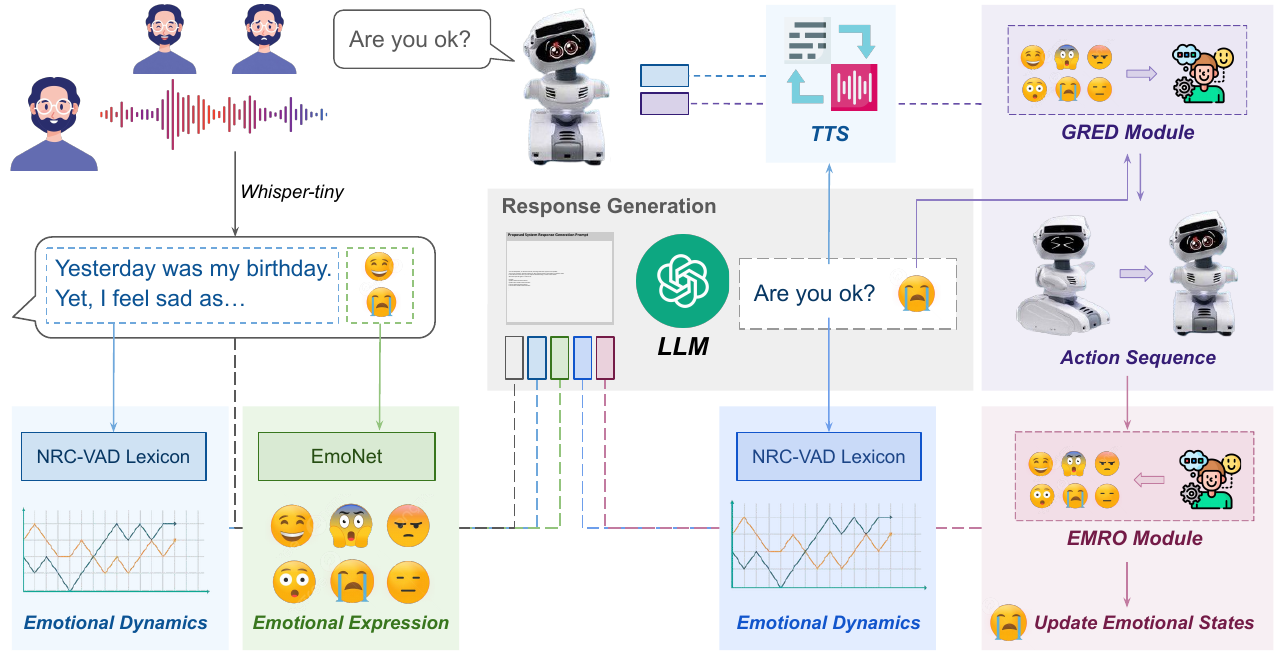} 
  \caption{Proposed \textsc{AffectLoop} system architecture in this study}
  \label{fig:architecture}
\end{figure*} 

\subsection{Emotional Dynamics and Speaker--Listener Alignment in Dialogue Systems} 

Many dialogue models treat emotion as a static label attached to an utterance or dialogue turn. However, emotion is dynamic: it changes over time, varies within and across utterances, and is influenced by the interaction partner. Emotion-dynamics research has studied how affective states evolve over time~\cite{ryan2025generative}, and dialogue-oriented work has shown that emotional shifts can be analyzed within conversational data~\cite{hipson2021emotion}. Recent LLM-based emotional-support research has also begun to move from snapshot-based evaluation to trajectory-based modeling. ETrajEval evaluates whether language models can improve and stabilize user emotional trajectories over time~\cite{tan2026detecting}, while AFlow models continuous affective flow along multi-turn emotional-support conversations~\cite{zou2026aflow}.

Beyond individual emotional trajectories, empathy also involves alignment between conversational partners. Prior psychological work shows that emotions can amplify speaker--listener neural alignment during communication~\cite{smirnov2019emotions}. This suggests that an empathetic system should not only respond to a user's current emotion, but also participate in the evolving affective exchange between speaker and listener. Existing dialogue studies rarely operationalize this idea in embodied systems. Our work addresses this gap by explicitly representing both the speaker's multimodal emotional dynamics and the listener's affective state.

\subsection{Affective and Multimodal Social Robots}


Social robots can express empathy through both verbal and nonverbal behaviors, including facial expressions, gestures, gaze, body motion, and speech style~\cite{pang2025human, pang2025does}. Prior studies have shown that nonverbal affective cues are important for human--robot interaction, particularly in social and emotionally supportive settings. With the development of LLMs, recent work has explored generating richer robot behaviors from dialogue context. SAFE uses an LLM to generate empathetic nonverbal cues for social robots, including speech, action, facial expression, and emotion~\cite{lee2023safe}. \cite{mishra2023realtime} use LLMs to generate real-time robot emotional displays during human--robot dialogue. Other work has explored multimodal emotional-support agents that incorporate visual or nonverbal context into response generation~\cite{schmidmaier2025nonverbal}.

These studies demonstrate the value of affective and multimodal behavior in social robots, but most focus either on recognizing user emotion or generating robot expressions. In contrast, our system connects these two sides through a speaker--listener affective loop: the robot conditions its response on the speaker's verbal and facial emotional dynamics while also incorporating its own verbal and behavioral affective state. This allows the robot to generate both spoken empathetic responses and emotionally congruent embodied behavior.

\section{Proposed System}

In this section, we describe the architecture of our proposed system, as shown
in Fig.~\ref{fig:architecture}. We implemented our system on the Misty II robot,
an open programmable robotics platform~\footnote{\url{https://www.mistyrobotics.com/misty-ii}}. The system is built as an incremental
multimodal spoken dialogue pipeline on Retico, a Python framework for
incremental spoken dialogue systems~\cite{michael2020retico}. It processes the
speaker's speech and facial behavior, estimates speaker- and listener-side
affective states, and uses these signals to guide LLM-based response generation
and robot behavior execution. The updated robot affective state is then fed
back into the next turn, forming a closed speaker--listener affective loop.

\subsection{Incremental Dialogue Framework}

Our system is implemented on top of Retico, a Python-based framework for building incremental spoken dialogue systems~\cite{michael2020retico}. Retico have been widely used previously on robot-ready spoken dialogue systems, child-level language interaction, and so on~\cite{kennington2020rrsds, levandovsky2025learning}. Retico follows the Incremental Unit (IU) model~\cite{schlangen2011general}, where each module processes and passes small units of information, such as audio frames, ASR hypotheses, images, or dialogue-state updates. This design allows the system to connect perception, language understanding, response generation, and robot behavior execution in a modular pipeline.

\subsection{Multimodal Input Module}

The system receives user input from two modalities: speech and facial behavior. For speech input, we use a hand microphone connected to a Retico audio stream. The audio is processed by an on-device ASR module based on Whisper-Tiny~\cite{radford2022whisper}, a lightweight ASR model for local incremental processing. The ASR module first buffers incoming audio frames and applies voice activity detection to estimate whether the user is currently speaking. During speech, partial recognition hypotheses are generated periodically. When silence is detected, the hypothesis is committed as the final user utterance and passed to the downstream dialogue modules.

For visual input, we use the robot camera to capture the user's facial behavior. We implement a Retico image module~\cite{manaseryan-etal-2025-rrsds} that produces image IUs from either a webcam, an IP camera stream, or a video source. Each image IU contains the current frame and frame-rate information, allowing the visual stream to be processed independently from the speech stream. 

\subsection{Affect Modeling Module}

We model affective states for both the human speaker and the robot listener.
For the speaker side, verbal affect is estimated from the ASR output using a
lexicon-based emotion tracking module. Instead of representing an utterance with
a single aggregate emotion score, the module incrementally computes a sequence
of valence, arousal, and dominance (VAD) vectors over the recognized tokens:
$\mathbf{e}^{s}_{1:n} = [\mathbf{e}^{s}_{1}, \ldots, \mathbf{e}^{s}_{n}]$,
where $\mathbf{e}^{s}_{t} = (v_t, a_t, d_t)$. These scores are computed using the NRC-VAD lexicon~\cite{mohammad2018vad}, a human-annotated lexicon providing valence, arousal, and dominance scores for over 20,000 English words, allowing the system to represent the speaker's verbal emotional dynamics within the current turn.

For nonverbal speaker affect, image frames captured by the robot camera are
processed by a facial expression recognition module. The module detects the
speaker's face and applies EmoNet~\cite{toisoul2021emonet}, a deep neural network designed for facial affect analysis under naturalistic conditions that jointly estimates categorical emotion, valence, and arousal. The frame-level estimates are aggregated over a short temporal window and used as the speaker's nonverbal emotional dynamics.

For the listener side, we estimate the robot's verbal affect from its generated
response using the same VAD-based tracking module, producing a listener verbal
VAD trajectory. The robot's nonverbal affect is estimated from its generated
action sequence using the EMRO action classifier~\cite{baral2025recognizing}, which maps robot behaviors to
six affective categories: \textit{anger/frustration},
\textit{confusion/sorrow/boredom}, \textit{disgust/surprise/alarm/fear},
\textit{interest/desire}, \textit{joy/hope}, and
\textit{understanding/gratitude/relief}. The listener verbal and nonverbal
affective states are stored and used as input for the next dialogue turn.

\begin{figure}[ht]
  \centering
  \includegraphics[width=0.85\linewidth]{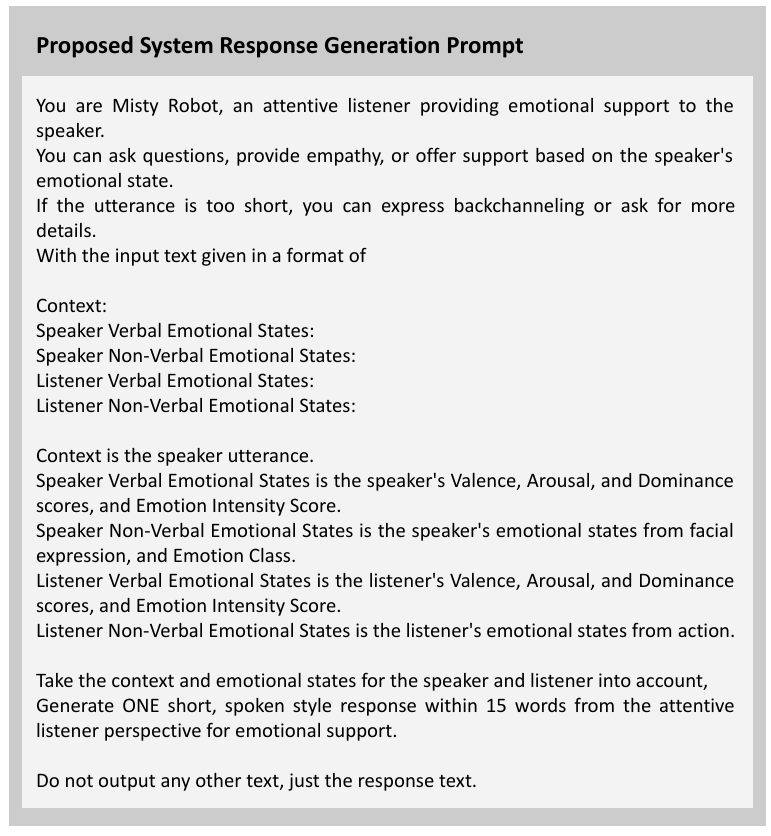} 
  \caption{Complete prompt template for the proposed speaker--listener
alignment empathetic response generation framework. The baseline used
the same prompt but excluded all speaker- and listener-side affective-state
information and the corresponding instruction to consider these states.}
  \label{fig:prompt}
\end{figure} 

\subsection{Response and Behavior Generation Module}

We use GPT-4.1-nano\footnote{\url{https://openai.com/index/gpt-4-1/}} as the LLM backbone for response generation, selected for interactive deployment. At each
turn, the system aggregates the dialogue context, the speaker's verbal VAD
trajectory, the speaker's facial affective trajectory, the listener's verbal VAD
trajectory, and the listener's nonverbal affective state. This speaker--listener
affective context is serialized into a structured prompt and provided to the LLM,
as shown in Fig.~\ref{fig:prompt}. The LLM then generates a short spoken
response from the perspective of an attentive listener.

In parallel, the LLM predicts an emotion label for the robot's next response.
This label is passed to the GRED action generation module~\cite{baral2025recognizing}, which generates a
sequence of robot behaviors conditioned on the predicted affective category. The
spoken response is synthesized with the robot's onboard text-to-speech engine, while the generated behavior
sequence is executed simultaneously. After execution, the robot's spoken
response and action sequence are analyzed again by the listener affect modules.
The updated listener affective state is then fed back into the next turn,
forming a closed speaker--listener affective loop.

\begin{figure}[ht]
  \centering
  \includegraphics[angle=-90,width=0.5\linewidth]{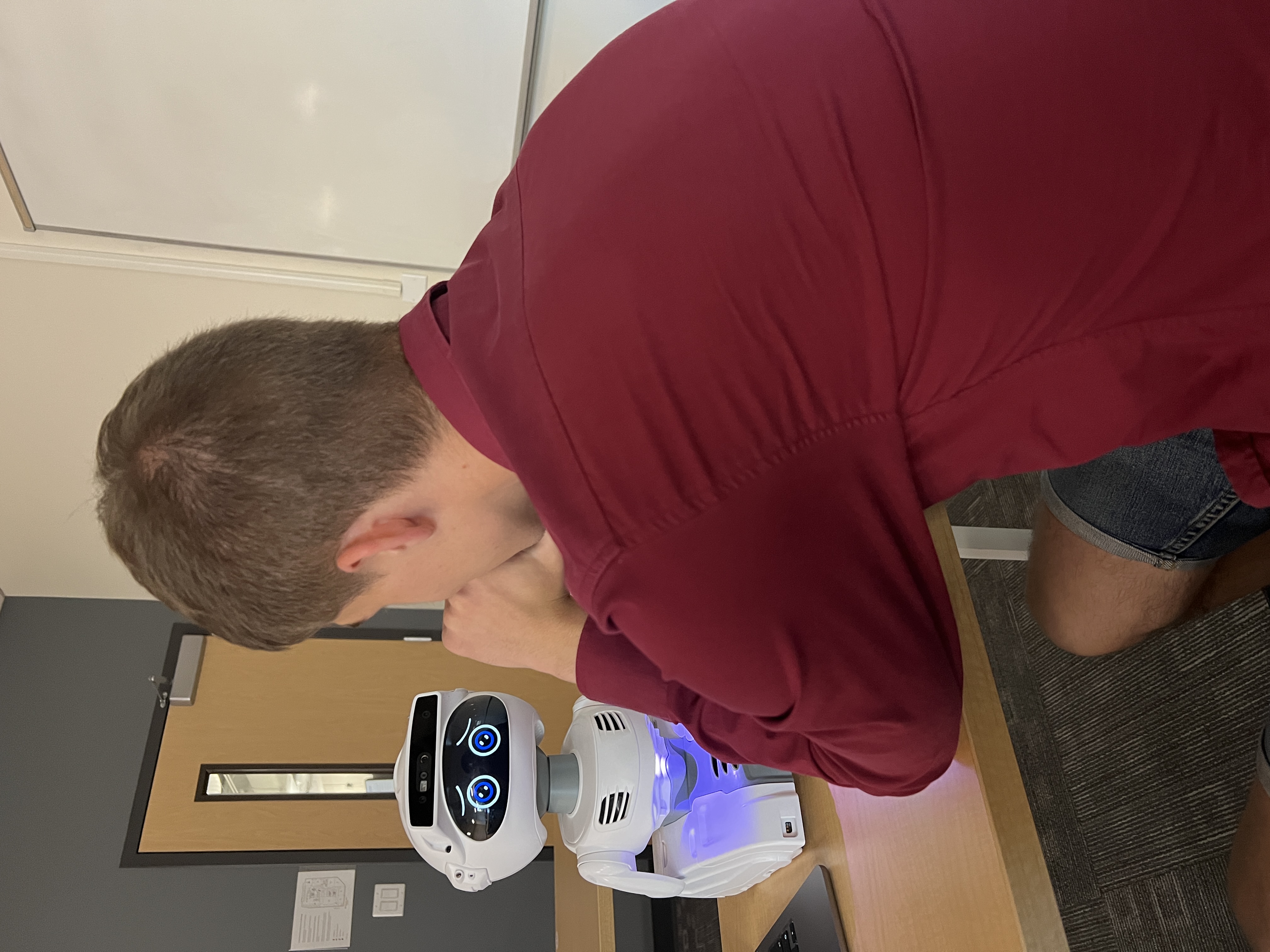} 
  \caption{
  Photo of interaction with Misty II by the participant
  }
  \label{fig:misty}
\end{figure} 

\section{Experimental Setup}

\subsection{Study Design}

We conducted a pilot within-subject user study to evaluate the proposed
speaker--listener emotion-dynamics-aware system. Each participant interacted
with two systems: a baseline system and the proposed system. The baseline
condition generated empathetic responses only from the speaker's utterance,
whereas the proposed condition incorporated both speaker-side and listener-side
dynamic affective states into the response-generation context.

Five participants took part in the study. All participants provided informed consent before participation. Each participant interacted with both
systems for five minutes, and the order of the two conditions was randomized to
reduce order effects. In both conditions, participants talked with the Misty II
robot in an open-ended attentive-listening setting. An example interaction is
shown in Figure~\ref{fig:misty}.

Following prior work evaluation metrics~\cite{charrier2019rope,kawai2024evaluation}, participants rated each
system using a 7-point Likert scale. The questionnaire covered four dimensions:
\textit{Naturalness}, \textit{Empathetic Listening},
\textit{Empathetic Response}, and \textit{User Satisfaction}. The detailed
questionnaire items are shown in Table~\ref{tab:eval_item}.

\subsection{Exploratory Interaction-Process Analysis}

In addition to subjective ratings, we conducted an exploratory post-hoc analysis
of the interaction logs to examine how the affective process differed between
the two conditions. We analyze two process-level metrics: dyadic affective alignment and speaker
affect shift.

\paragraph{Dyadic affective alignment.}
Let $\mathbf{s}_{t} = (v^s_t, a^s_t, d^s_t)$ denote the speaker VAD vector at
turn $t$. Let $\mathbf{l}^{pre}_{t}$ and $\mathbf{l}^{post}_{t}$ denote the
robot listener VAD state before and after generating its response. We define
the listener affective change at turn $t$ as: $\Delta \mathbf{l}_{t} = \mathbf{l}^{post}_{t} - \mathbf{l}^{pre}_{t}.$ We then measure how closely the listener's affective change aligns with the
speaker's affective state by computing cosine similarity:
\[
    A_t =
    \cos(\mathbf{s}_{t}, \Delta \mathbf{l}_{t})
    =
    \frac{\mathbf{s}_{t} \cdot \Delta \mathbf{l}_{t}}
    {\|\mathbf{s}_{t}\| \|\Delta \mathbf{l}_{t}\|}.
\]
Turns for which either vector has zero norm are excluded from this metric. We
also report a normalized alignment score:
\[
    A^{norm}_t = \frac{A_t + 1}{2},
\]
which maps cosine similarity from $[-1,1]$ to $[0,1]$.

To analyze alignment at the level of individual affective dimensions, we
compute sign-match rates for valence, arousal, and dominance. For each
dimension $k \in \{v,a,d\}$, sign match is defined as:
\[
    M^k_t =
    \mathbb{I}
    \left[
    \operatorname{sign}(s^k_t)
    =
    \operatorname{sign}(\Delta l^k_t)
    \right],
\]
where turns with near-zero values in either term are excluded. The sign-match
rate is the mean of $M^k_t$ over valid turns. Intuitively, this metric measures
whether the robot listener's affective movement follows the same directional
tendency as the speaker's affective state in each VAD dimension.

\paragraph{Speaker affect shift and distress recovery.}
To examine how the speaker's affective state changed after the robot response,
we computed affect shifts between adjacent speaker turns. For consecutive
speaker VAD states $\mathbf{s}_{t}$ and $\mathbf{s}_{t+1}$, we define: $\Delta \mathbf{s}_{t} = \mathbf{s}_{t+1} - \mathbf{s}_{t}.$
We report the mean change in valence, arousal, and dominance across all valid
adjacent speaker-turn transitions.

We further compute a valence-based distress recovery score for turns in which
the speaker's current valence is negative. Let $v_t$ and $v_{t+1}$ denote the
speaker valence at consecutive turns. Distress recovery is defined as:
\[
    R_t =
    |\min(v_t, 0)| - |\min(v_{t+1}, 0)|.
\]
A positive value indicates that the magnitude of negative valence decreased in
the next speaker turn, suggesting recovery from a negative affective state. A
negative value indicates that negative valence increased. We report the mean
distress recovery score over turns with $v_t < 0$, as well as the positive
recovery rate:
\[
    P_{rec}
    =
    \frac{1}{|\mathcal{T}_{neg}|}
    \sum_{t \in \mathcal{T}_{neg}}
    \mathbb{I}[R_t > 0],
\]
where $\mathcal{T}_{neg} = \{t \mid v_t < 0\}$.

\begin{table}[h]
\caption{Impression Ratings Items}
\label{tab:eval_item}
\centering

\begin{threeparttable}
\begin{tabularx}{\linewidth}{lYcc}
\toprule
\multicolumn{1}{c}{Item}
& \multicolumn{1}{c}{Description}
& Baseline
& Proposed \\
\hline

\rowcolor[gray]{0.95}
\multicolumn{4}{l}{\textbf{Naturalness}} \\
Q1  & The robot's responses were human-like. & 5.00 & 4.80\\
Q2  & The words the robot used were natural. & 4.80 & 5.20\\
Q3  & The robot's responses could stimulate my own talk. & 4.00 & 4.60\\
Q4  & The robot understood my talk. & 5.40 & 5.20\\
Q5  & I believe the robot was fully autonomous
(i.e., not controlled by a human behind the scenes). & 4.00 & 3.80\\
\hline
\multicolumn{2}{c}{\textit{Average}} & 4.80 & 4.95 \\

\midrule
\rowcolor[gray]{0.95}
\multicolumn{4}{l}{\textbf{Empathetic Listening}} \\
Q6 & The robot displayed empathy towards me. & 5.60 & 5.60\\
Q7 & The robot took the conversation seriously. & 5.20 & 5.60\\
Q8 & The robot was listening intently. & 5.20 & 5.40\\
Q9 & The robot was listening actively. & 5.20 & 5.00\\
Q10 & The robot showed interest in the conversation. & 4.80 & 5.40\\
\hline
\multicolumn{2}{c}{\textit{Average}} & 5.20 & 5.40 \\

\midrule
\rowcolor[gray]{0.95}
\multicolumn{4}{l}{\textbf{Empathetic Response}} \\
Q11 & The robot comforts me when I am upset. & 5.20 & 5.40\\
Q12 & The robot encourages me. & 5.40 & 6.20\\
Q13 & The robot praises me when I have done something well. & 4.40 & 5.40\\
Q14 & The robot helps me when I need it. & 4.20 & 4.40\\
\hline
\multicolumn{2}{c}{\textit{Average}} & 4.80 & 5.35 \\

\midrule
\rowcolor[gray]{0.95}
\multicolumn{4}{l}{\textbf{User Satisfaction}} \\
Q15 & The robot was easy to talk to. & 4.80 & 5.00\\
Q16 & I want to talk with the robot again. & 4.60 & 4.80\\
Q17 & The conversation was smooth. & 3.00 & 3.80\\
Q18 & I was satisfied with the conversation. & 4.60 & 5.00\\
Q19 & After the conversation, I felt better
(my stress/negative emotions were reduced). & 4.00 & 4.80\\
Q20 & I felt anxious when interacting with the robot. & 3.00 & 2.80\\
\hline
\multicolumn{2}{c}{\textit{Average}} & 4.20 & 4.68 \\

\midrule
\multicolumn{2}{c}{\textbf{\textit{Overall Average}}}
& 4.75 & 5.10 \\
\bottomrule
\end{tabularx}

\begin{tablenotes}[flushleft]
\footnotesize
\item \textbf{Note:} Q5 and Q20 were excluded from the average calculation, as Q5 serves as a perceived-autonomy check and Q20 measures interaction anxiety rather than user satisfaction.
\end{tablenotes}

\end{threeparttable}
\end{table}

\section{Results and Discussion}

Table~\ref{tab:eval_item} shows that the proposed system received higher
overall impression ratings than the baseline, increasing from 4.75 to 5.10.
The largest gain appeared in \textit{empathetic response} (4.80 to 5.35),
suggesting that the proposed speaker--listener affective conditioning mainly
improved how supportive the robot's responses felt. This trend is also reflected
in individual items: the proposed system was rated higher for encouraging the
user, praising the user, comforting the user, and helping when needed. User
satisfaction also improved from 4.20 to 4.68, including higher ratings for
conversation smoothness, satisfaction, and feeling better after the
conversation. Naturalness and empathetic listening showed smaller but positive
average gains. However, the proposed system was not uniformly better on every
item; the baseline was slightly higher for perceived human-likeness,
understanding the user's talk, and active listening. These mixed item-level
results suggest that the proposed method improved perceived emotional support
more clearly than general conversational naturalness.

To further examine whether these subjective trends were reflected in the
interaction process, we conducted an exploratory post-hoc analysis of the
affective trajectories in the logs. Table~\ref{tab:posthoc_process} summarizes
the metrics introduced in the previous subsection. The proposed system showed
higher speaker--listener affective alignment, with mean cosine alignment
$\bar{A}$ increasing from 0.053 to 0.169 and normalized alignment
$\bar{A}^{norm}$ increasing from 0.526 to 0.585. The clearest dimensional
change was observed in valence sign match $M^v$, which increased from 46.4\%
to 59.7\%, while arousal sign match $M^a$ remained similar. This suggests that
the proposed system better followed the user's positive--negative affective
direction, rather than simply increasing synchrony across all affective
dimensions.

The proposed system also showed stronger valence-based distress recovery.
Mean recovery $\bar{R}$ increased from 0.151 to 0.339, and the positive
recovery rate $P_{rec}$ increased from 72.2\% to 100.0\%. This indicates that
when the user's valence was negative, the next user turn was more often less
negative after interacting with the proposed system. Together, the subjective
ratings and post-hoc process analysis suggest a consistent trend: conditioning
the LLM on both speaker emotional dynamics and listener affective state may
help close the affective loop in embodied empathetic interaction.

\begin{table}[t]
\centering
\small
\caption{Post-hoc interaction-process analysis [\%]. Differences from the baseline are shown beside the proposed values.}
\label{tab:posthoc_process}
\setlength{\tabcolsep}{3pt}
\begin{tabular}{@{}l@{\hspace{8pt}}r@{\hspace{8pt}}r@{}}
\toprule
\multicolumn{1}{c}{Metric} &
\multicolumn{1}{c}{Baseline} &
\multicolumn{1}{c}{Proposed} \\
\midrule
\multicolumn{3}{@{}l}{\textbf{Speaker--listener alignment}} \\
$\bar{A}$: mean cosine alignment
& 5.29
& 16.92 \, \textcolor{ForestGreen}{$\uparrow\!11.64$} \\
$\bar{A}^{norm}$: mean normalized alignment
& 52.64
& 58.46 \, \textcolor{ForestGreen}{$\uparrow\!5.82$} \\
$M^v$: valence sign match
& 46.41
& 59.67 \, \textcolor{ForestGreen}{$\uparrow\!13.26$} \\
$M^a$: arousal sign match
& 54.59
& 53.67 \, \textcolor{BrickRed}{$\downarrow\!0.92$} \\
$M^d$: dominance sign match
& 54.72
& 57.42 \, \textcolor{ForestGreen}{$\uparrow\!2.71$} \\
\midrule
\multicolumn{3}{@{}l}{\textbf{Speaker affect shift}} \\
$\overline{\Delta v^s}$: mean valence shift
& 0.65
& -2.49 \, \textcolor{BrickRed}{$\downarrow\!3.14$} \\
$\overline{\Delta a^s}$: mean arousal shift
& -0.84
& -0.43 \, \textcolor{ForestGreen}{$\uparrow\!0.41$} \\
$\overline{\Delta d^s}$: mean dominance shift
& -0.49
& 0.80 \, \textcolor{ForestGreen}{$\uparrow\!1.29$} \\
\midrule
\multicolumn{3}{@{}l}{\textbf{Distress recovery}} \\
$\bar{R}$: mean distress recovery
& 15.15
& 33.90 \, \textcolor{ForestGreen}{$\uparrow\!18.75$} \\
$P_{rec}$: positive recovery rate
& 72.22
& 100.00 \, \textcolor{ForestGreen}{$\uparrow\!27.78$} \\
\bottomrule
\end{tabular}
\end{table}

\section{Conclusion}

We presented \textsc{AffectLoop}, a multimodal
speaker--listener emotion-dynamics-aware spoken dialogue system for an
empathetic social robot. Unlike text-only empathetic dialogue systems, our
system conditions LLM-based response generation on both the speaker's verbal and
facial emotional dynamics and the robot listener's own verbal and behavioral
affective state. This enables the robot to generate not only spoken empathetic
responses, but also emotionally congruent embodied behavior within a closed
affective loop.

A pilot within-subject study showed that the proposed system was rated higher
than a text-only LLM baseline in overall impression, with the clearest gains in
empathetic response and user satisfaction. The post-hoc interaction-log
analysis further suggested that the proposed system produced stronger
speaker--listener affective alignment and greater valence-based distress
recovery. These findings provide preliminary evidence that incorporating both
speaker emotional dynamics and listener affective state can make social-robot
responses feel more supportive and can positively shape the affective process of
the interaction.

\section*{Acknowledgment}

This material was based upon work supported by the National Science Foundation under Grant No. 2343118 and JST Moonshot R\&D JPMJPS2011.










\printbibliography

\end{document}